\documentstyle[prd,aps,12pt,epsf]{revtex}
\begin{document}

\baselineskip=7mm
\def\ap#1#2#3{           {\it Ann. Phys. (NY) }{\bf #1} (19#2) #3}
\def\arnps#1#2#3{        {\it Ann. Rev. Nucl. Part. Sci. }{\bf #1} (19#2) #3}
\def\cnpp#1#2#3{        {\it Comm. Nucl. Part. Phys. }{\bf #1} (19#2) #3}
\def\apj#1#2#3{          {\it Astrophys. J. }{\bf #1} (19#2) #3}
\def\asr#1#2#3{          {\it Astrophys. Space Rev. }{\bf #1} (19#2) #3}
\def\ass#1#2#3{          {\it Astrophys. Space Sci. }{\bf #1} (19#2) #3}

\def\apjl#1#2#3{         {\it Astrophys. J. Lett. }{\bf #1} (19#2) #3}
\def\ass#1#2#3{          {\it Astrophys. Space Sci. }{\bf #1} (19#2) #3}
\def\jel#1#2#3{         {\it Journal Europhys. Lett. }{\bf #1} (19#2) #3}

\def\ib#1#2#3{           {\it ibid. }{\bf #1} (19#2) #3}
\def\nat#1#2#3{          {\it Nature }{\bf #1} (19#2) #3}
\def\nps#1#2#3{          {\it Nucl. Phys. B (Proc. Suppl.) } {\bf #1} (19#2) #3} 
\def\np#1#2#3{           {\it Nucl. Phys. }{\bf #1} (19#2) #3}

\def\pl#1#2#3{           {\it Phys. Lett. }{\bf #1} (19#2) #3}
\def\pr#1#2#3{           {\it Phys. Rev. }{\bf #1} (19#2) #3}
\def\prep#1#2#3{         {\it Phys. Rep. }{\bf #1} (19#2) #3}
\def\prl#1#2#3{          {\it Phys. Rev. Lett. }{\bf #1} (19#2) #3}
\def\pw#1#2#3{          {\it Particle World }{\bf #1} (19#2) #3}
\def\ptp#1#2#3{          {\it Prog. Theor. Phys. }{\bf #1} (19#2) #3}
\def\jppnp#1#2#3{         {\it J. Prog. Part. Nucl. Phys. }{\bf #1} (19#2) #3}

\def\rpp#1#2#3{         {\it Rep. on Prog. in Phys. }{\bf #1} (19#2) #3}
\def\ptps#1#2#3{         {\it Prog. Theor. Phys. Suppl. }{\bf #1} (19#2) #3}
\def\rmp#1#2#3{          {\it Rev. Mod. Phys. }{\bf #1} (19#2) #3}
\def\zp#1#2#3{           {\it Zeit. fur Physik }{\bf #1} (19#2) #3}
\def\fp#1#2#3{           {\it Fortschr. Phys. }{\bf #1} (19#2) #3}
\def\Zp#1#2#3{           {\it Z. Physik }{\bf #1} (19#2) #3}
\def\Sci#1#2#3{          {\it Science }{\bf #1} (19#2) #3}

\def\n.c.#1#2#3{         {\it Nuovo Cim. }{\bf #1} (19#2) #3}
\def\r.n.c.#1#2#3{       {\it Riv. del Nuovo Cim. }{\bf #1} (19#2) #3}
\def\sjnp#1#2#3{         {\it Sov. J. Nucl. Phys. }{\bf #1} (19#2) #3}
\def\yf#1#2#3{           {\it Yad. Fiz. }{\bf #1} (19#2) #3}
\def\zetf#1#2#3{         {\it Z. Eksp. Teor. Fiz. }{\bf #1} (19#2) #3}
\def\zetfpr#1#2#3{         {\it Z. Eksp. Teor. Fiz. Pisma. Red. }{\bf #1} (19#2) #3}
\def\jetp#1#2#3{         {\it JETP }{\bf #1} (19#2) #3}
\def\mpl#1#2#3{          {\it Mod. Phys. Lett. }{\bf #1} (19#2) #3}
\def\ufn#1#2#3{          {\it Usp. Fiz. Naut. }{\bf #1} (19#2) #3}
\def\sp#1#2#3{           {\it Sov. Phys.-Usp.}{\bf #1} (19#2) #3}
\def\ppnp#1#2#3{           {\it Prog. Part. Nucl. Phys. }{\bf #1} (19#2) #3}
\def\cnpp#1#2#3{           {\it Comm. Nucl. Part. Phys. }{\bf #1} (19#2) #3}
\def\ijmp#1#2#3{           {\it Int. J. Mod. Phys. }{\bf #1} (19#2) #3}
\def\ic#1#2#3{           {\it Investigaci\'on y Ciencia }{\bf #1} (19#2) #3}
\def\tp{these proceedings}
\def\pc{private communication}
\def\ip{in preparation}
\relax
\newcommand{\TeV}{\,{\rm TeV}}
\newcommand{\GeV}{\,{\rm GeV}}
\newcommand{\MeV}{\,{\rm MeV}}
\newcommand{\keV}{\,{\rm keV}}
\newcommand{\eV}{\,{\rm eV}}
\newcommand{\Tr}{{\rm Tr}\!}
\renewcommand{\arraystretch}{1.2}
\newcommand{\be}{\begin{equation}}
\newcommand{\ee}{\end{equation}}
\newcommand{\bea}{\begin{eqnarray}}
\newcommand{\eea}{\end{eqnarray}}
\newcommand{\ba}{\begin{array}}
\newcommand{\ea}{\end{array}}
\newcommand{\bmat}{\left(\ba}
\newcommand{\emat}{\ea\right)}
\newcommand{\refs}[1]{(\ref{#1})}
\newcommand{\ler}{\stackrel{\scriptstyle <}{\scriptstyle\sim}}
 \newcommand{\ger}{\stackrel{\scriptstyle >}{\scriptstyle\sim}}
\newcommand{\lag}{\langle}
\newcommand{\rag}{\rangle}
\newcommand{\ns}{\normalsize}
\newcommand{\cm}{{\cal M}}
\newcommand{\gr}{m_{3/2}}
\newcommand{\p}{\partial}
\def\321{$SU(3)\times SU(2)\times U(1)$}
\def\tl{{\tilde{l}}}
\def\tL{{\tilde{L}}}
\def\bd{{\overline{d}}}
\def\tL{{\tilde{L}}}
\def\a{\alpha}
\def\b{\beta}
\def\g{\gamma}
\def\c{\chi}
\def\d{\delta}
\def\D{\Delta}
\def\db{{\overline{\delta}}}
\def\Db{{\overline{\Delta}}}
\def\e{\epsilon}
\def\l{\lambda}
\def\n{\nu}
\def\m{\mu}
\def\nt{{\tilde{\nu}}}
\def\p{\phi}
\def\P{\Phi}
\def\x{\xi}
\def\r{\rho}
\def\s{\sigma}
\def\t{\tau}
\def\th{\theta}
\def\ne{\nu_e}
\def\nm{\nu_{\mu}}
\def\snui{\tilde{\nu_i}}
\renewcommand{\Huge}{\Large}
\renewcommand{\LARGE}{\Large}
\renewcommand{\Large}{\large}
\title{
\hfill hep-ph/9808232\\
\hfill PRL-TH- 98/008\\
Gauge mediated supersymmetry breaking and neutrino anomalies} 

\author{ Anjan S. Joshipura and Sudhir K. Vempati \\
{\ns\it Theoretical Physics Group, Physical Research Laboratory,}\\
{\ns\it Navarangpura, Ahmedabad, 380 009, India.}}
\maketitle
\begin{center}
{\bf Abstract}
\end{center}
\begin{abstract}
Supersymmetric standard model with softly broken lepton symmetry provides
a suitable framework to accommodate the solar and atmospheric neutrino
anomalies. This model contains a natural explanation for large mixing
and hierarchal masses without fine tuning of the parameters. Neutrino
spectrum is particularly constrained in the minimal messenger model (MMM)
of gauge mediated SUSY breaking, since all SUSY breaking effects are
controlled in MMM by a single parameter. We study the structure of 
neutrino masses and mixing both in MMM and in simple extensions of it
in the context of solar and atmospheric neutrino anomalies.
\end{abstract}

\section{Introduction}

The hints in favour of a non-zero neutrino mass are greatly strengthened
by the recent high statistics results of the atmospheric $\nu_\m$ deficit 
at the Superkamioka \cite{sk}. These results are claimed to be the
evidence of a
non-zero neutrino (mass)$^2$ difference $\Delta_A\sim (.07 \eV)^2$
and large mixing between $\n_\m$ and $\n_\t$ or a sterile state.
Additional hint for one more mass scale comes from the observed deficit in
the solar neutrinos. These results require \cite{bsk} the solar mass scale
$\Delta_S$
of $10^{-5}-10^{-6} \eV^2$ (MSW conversion \cite{msw} ) or 
$10^{-10} \eV^2$ (vacuum
oscillations). While the former can be consistent with the small mixing
of $\n_e$, the latter explanation requires \cite {bsk} one more large mixing. 
On the basis of these results, the neutrino spectrum seems to be
characterized by hierarchical masses and one or two large mixings.
The above neutrino spectrum
is quite different from the quark spectrum  suggesting a
characteristically different origin for the neutrino masses and mixing. 

The standard seesaw model \cite{lang} based on simple $SO(10)$ theory 
links respectively the $\nu$-masses and leptonic mixing to the up 
quark masses and the KM matrix.  It leads to
${\Delta_S\over \Delta_A}\sim \left( {m_c\over m_t}\right)^4\sim 10^{-8}$
if simplifying assumptions are made on the structure of the right handed
neutrino masses \cite {lang}. With $\Delta_A\sim 5\times 10^{-3} \eV^2$
 one obtains $\Delta_S\sim 5 \times 10^{-11}\eV^2$ in the range required for
 the vacuum solution to the solar anomalies. The MSW and the
atmospheric mass scales cannot be easily reconciled in this simple
 picture although additional contribution from a left handed triplet 
\cite{triplet} field can be used to do so. Moreover, large mixing angles 
though not impossible \cite{smir} are not natural in the seesaw picture. 

 We wish to discuss an alternative scheme for neutrino masses based on
supersymmetry. The scheme utilizes soft lepton number violation and
soft supersymmetry (SUSY) breaking terms generated by gauge mediated
interactions.
As we will see, this scheme  is quite predictive and can
provide
natural and simultaneous understanding of the hierarchical masses
and large mixings in the leptonic sector hinted on experimental grounds.

The Minimal Supersymmetric Standard Model (MSSM) naturally contains lepton
number violation
if the conventional R symmetry is not imposed on it. This violation is 
manifested through hard (trilinear) or soft (bilinear) operators \cite{hs}.
The soft
terms are parameterized in terms of three dimensionful co-efficients
$\epsilon_i$:

\be \label {bi}
W = \epsilon_i L'_i H_2
\ee
 
We will assume that eq (\ref {bi}) is the only source of lepton
number violation. This
assumption is theoretically well motivated. Spontaneous breaking of
lepton number \cite{masvalle} could normally result in such a term.
Alternatively, one could imagine a generalized
Peccei-Quinn symmetry whose spontaneous breaking leads to $\mu$ and 
$\epsilon_i$ at the weak scale through dim 5 operators
\cite{tamvakis,asb}.
Moreover, it is 
possible to choose the PQ charges of different fields in such a way that
the generation of effective trilinear operators is enormously suppressed.

The presence  of $\epsilon_i$ leads to neutrino masses and mixing among 
them. In fact, these three parameters control three neutrino masses and
three mixing angles. This is easily seen from the fact that the limit
$\epsilon_i \rightarrow 0$ corresponds to no lepton number violation, zero
neutrino masses and trivial mixing in the leptonic sector. In practice,
the neutrino masses also depend upon parameters responsible for the SUSY
breaking. But this breaking can be characterized by only one parameter
in a minimal version of the gauge mediated breaking \cite{gm1} of SUSY.
This scenario
can therefore provide a constrained framework for the description of the 
neutrino masses and mixing in which four input parameters determine three
masses and three mixing angles.

The neutrino spectrum resulting from eq.(1) has been extensively studied 
in the context of MSSM with supergravity induced SUSY breaking
\cite{asb,numass}. The following
features  make it a very attractive scenario for the description of the
neutrino spectrum:

(i) Neutrinos obtain masses \cite{hs} through two sources: through mixing with
gauginos induced by sneutrino vacuum expectation value (vev) and through
coupling of neutrinos
to squarks and sleptons. Both these contributions are suppressed by the
$b$ and $\tau$ Yukawa couplings in the supergravity framework as well as
in the gauge mediated models. This leads to suppressed ($\sim$ MeV)
neutrino masses even when $\epsilon_i$ are large ($\sim$ 100 GeV).

(ii) The sneutrino vev makes only one combination of neutrinos massive.
The other combinations pick up masses from the loop diagrams \cite{loop}.
In particular,
one of the neutrinos remains massless when Yukawa couplings of the first two
generations are neglected \cite{asb}. This generates clear hierarchy in
masses of all
the three neutrinos without fine tuning or without imposing any horizontal
symmetries.

(iii) Mixing among neutrinos is essentially controlled \cite{asm} by
ratios of 
$\epsilon_i$ which can be large when these parameters are not
hierarchical. One can therefore simultaneously obtain
large mixing and hierarchal masses. 

The above features (ii) and (iii) can allow simultaneous solution of the
solar and atmospheric neutrino anomalies.

We discuss the details of this in the following section. Section (2)
summarizes the analytic structure of neutrino masses and mixing one obtains
in the presence of eq.(1). In the next section, we summarize salient
features of the gauge mediated models and introduce the minimal
model in this category. Section (4) contains detailed prediction of this
model and phenomenological discussions on the solar and the atmospheric
neutrino problem. We introduce an extended version of the minimal model
which is
capable of solving the solar and atmospheric neutrino anomalies
simultaneously in Section (5). We end with a discussion in the last section.
\section{Neutrino masses and SUSY}
The structure of neutrino masses crucially depends
upon the nature of SUSY breaking soft terms. It is now recognized 
\cite {asb,numass} that neutrino masses are calculable in terms of 
basic parameters if soft terms associated with the leptonic and 
one of the Higgs doublet
$(H_1')$ superfields are identical at some high scale $\Lambda$. This 
happens in the minimal supergravity model as well as in gauge mediated
models. The latter class of models contain fewer parameters and thus are
more predictive. We will specialize to this case in the next section. 
The detailed analytic discussion of the structure of neutrino masses in the
present context was given in \cite{asb}. Here we briefly
recall some salient features to be used in this paper. Consider
the full superpotential of the MSSM \cite{fn1} including the
lepton number violating term ~(\ref{bi}):
\be \label{sup}
W= \l_{i}L_{i}'E_{i}^c H_1'+\m'H_1'H_2+\e_{i}L_{i}'H_2+\l_{i}^D
Q_{i}D^c_{i}H_1'+\l_{ij}^U
Q_{i}U^c_{j}H_2 \;, \ee
where,  $ i,j=1,2,3.$ We have made a specific choice of the basis such that 
$L_i' (Q_i)$ denote the mass eigenstates of the charged lepton (down
quarks) in the absence of the $\e_i$ terms. 
Following \cite{hs} we now
redefine the leptonic and the Higgs fields in such a way that 
the superpotential (\ref{sup})
does not contain bilinear $\e_i$-dependent terms.
This happens in the following unprimed basis:  
\bea \label {h1}
H_1&=&c_3 H_1'+s_3(s_2(c_1L_2'+s_1L_1')+c_2L_3'), \nonumber \\
L_1&=&-s_1L_2'+c_1L_1', \nonumber \\   
L_2&=&c_2(c_1L_2'+s_1L_1')-s_2L_3',  \nonumber \\  
L_3&=&-s_3 H_1'+c_3(s_2(c_1L_2'+s_1L_1')+c_2L_3'). 
\eea
Where,

\be \label{angles} \ba{cc}
\e_1=\m s_1s_2s_3,\;\;\;\;&\e_2=\m c_1s_2s_3,\\
\e_3=\m s_3c_2,\;\;\;&\m'=\m c_3, \ea \ee 
and  $\m\equiv 
(\m'^2+\e_1^2+\e_2^2+\e_3^2)^{1/2}$.

The superpotential in the new basis contains trilinear lepton number
violating terms. Moreover, the originally diagonal charged lepton mass
matrix $M_l$ now acquires \cite{asb} non-diagonal parts\cite{f2}:
\be \label{ml}
{M_l\over <H_1>}= \left (
\begin{array}{ccc} \l_1c_1c_3&-\l_2 s_1c_3&0\\
                  \l_1s_1c_2c_3&\l_2 c_1c_2c_3&-\l_3s_2c_3\\
                  \l_1s_1s_2&\l_2c_1s_2&\l_3c_2 \end{array} \right ) \ee
The mass basis for the charged leptons in the presence of non-zero $\e_i$
are denoted by $\a=e,\mu,\tau$ and are defined as:
\be \label{weak} 
L_i=(O_L^T)_{i\a} L_{\a},\;\;\;\; e^c_i=(O_R^T)_{i\a} e_{\a}^c, \ee
where,
\be\label{dia} O_LM_lO_R^T=diagonal. \ee

Note that the parameters $\l_i$ which denote charged lepton masses when
$\e_i=0$ still need to be hierarchical if $M_l$ in (\ref{ml}) is to
reproduce the charged lepton masses. One can determine \cite{asb} $O_L$ by
assuming
$\l_1\ll \l_2\ll \l_3$ and neglecting $\l_1$:
\be \label{ol} 
O_L^{T}\approx N_1\left( 
 \ba{ccc}                c_1&-s_1N_2&0\\
                 s_1 c_2&c_1 c_2N_2^{-1}&-s_2c_3N_1^{-1}N_2^{-1}\\
                s_1 s_2c_3&c_1s_2c_3N_2^{-1}&c_2N_1^{-1}N_2^{-1}\\ \ea
\right) 
\left( 
 \ba{ccc}                1&0&0\\
                 0&\cos\theta_{23}&-\sin\theta_{23}\\
                0&\sin\theta_{23}&\cos\theta_{23}\\ \ea
\right). \ee
Where,
\bea
N_1&=&(1-s_1^2s_2^2s_3^2)^{(-1/2)}, \nonumber \\
N_2&=&(1-s_2^2s_3^2)^{(1/2)}, \nonumber \\
\theta_{23}&\approx&N_1 c_1 c_2 s_2 \left( {s_3
m_{\mu}\over c_3m_{\tau}}\right )^2 .
\eea 
$\theta_{23}$ is small even for $s_{1,2,3}\sim O(1)$.
 We shall therefore neglect it.

The trilinear terms generated in the superpotential due to rotation
 (\ref{h1}) assume the following form \cite{asb} in the physical
mass eigenstate basis of charged leptons:
\be \label{w3}
W= -{\tan\theta_3\over <H_1>}[(O_L^T)_{3\a}L_{\a}]\left(
 m^l_{\b}L_{\b}e^c_{\b}+m_i^DQ_id_i^c \right).
\ee

The above trilinear terms generated due to the basis rotation lead to
neutrino masses at 1-loop \cite{loop}. The other contribution to neutrino
mass is generated by the soft SUSY breaking terms in a manner 
discussed below.

\subsection{Sneutrino vevs and neutrino masses }

The soft supersymmetry breaking part of the scalar potential contains 
linear terms in sneutrino fields and lead to their vev. 
These follow from the following soft terms written in the primed basis:
\begin{equation}\label{soft}\begin{array}{ccc}
V_{soft}&=&
m_{H_1^{\prime
2}}|H^{0\prime}_1|^2+m_{H_2^2}|H_2^0|^2+m_{\tilde{\nu}_i^{\prime^2}}  
|\tilde{\nu}_i^{\prime}|^2 \nonumber \\ 
&-&  \left[ H_2^0\left( \mu'B_{\m} H_1^{\prime 0}+\e_i
B_i\tilde{\nu'}_i\right )\;\; + H. c.\right]. \end{array}  
\end{equation}

It is always possible to choose a minimum with zero sneutrino vev if 
$B_\m=B_i$ and $m_{H_1^{\prime
2}}=m_{\n_i'}^2$. But, these equalities are generally not satisfied at the
weak scale
even if they are imposed at a high scale. This is due to
the presence of Yukawa couplings
which distinguish between Higgs and leptons. If one neglects the Yukawa
couplings of the first two generations then the soft mass parameters
related to the first two leptonic generations evolve in the same way.
Thus, one would have the following non-zero differences among low energy
parameters:
\be \label{deltas1} 
\Delta m_{L,H}\equiv (m_{\tilde{\nu}_3^{\prime^2}}
-m_{\tilde{\nu}_2',H_1'}^2)~,\;\;\;\; 
\Delta B_{L,H}\equiv (B_3-B_{2 ,\mu}). \ee
These differences determine the sneutrino vev. The latter are obtained by
minimizing the scalar potential expressed in the redefined basis of
eqs.(\ref {h1}). One finds:
\bea \label{vfull}
V&=&(m_{H_1}^2+\m^2)|H_1^0|^2+(m_{H_2}^2+\m^2)|H_2^0|^2+m_{\tilde{\nu}_3}^2
|\tilde{\nu_3}|^2
+ m_{\tilde{\nu}_2'}^2 |\tilde{\nu_2}|^2 + m_{\tilde{\nu_1}'}^2
|\tilde{\nu_1}|^2 \nonumber\\ 
&-& \left[ \mu H_2^0\left( B H_1^0+ c_3s_3 \tilde{\nu}_3( \Delta
B_H-s_2^2 
\Delta B_L)
- c_2 s_2 s_3 \tilde{\nu}_2  \Delta B_L\right)\;\; + H. c.\right ]
\nonumber \\
&+&\left[ -c_2 s_2 \Delta m_L (s_3H_1^0+ c_3 \tilde{\nu}_3)
\tilde{\nu}_2^*
+ s_3c_3 \tilde{\nu}_3 H_1^{0*} ( \Delta m_H- s_2^2 \Delta m_L)
\;\;+\;H.c.
\right] \nonumber\\ 
&+& {1\over 8} (g^2+g'^2) ( |H_1^0|^2+|\tilde{\nu}_1|^2
+|\tilde{\nu}_2|^2+|\tilde{\nu}_3|^2-|H_2^0|^2)^2 .
\eea
Where, 
\bea \label{vfull1}
m_{H_1}^2&=&m_{H_1'}^2+s_3^2\Delta m_H-s_2^2 s_3^2 \Delta m_L,
 \nonumber \\ 
m_{\tilde{\nu}_3}^2&=&m_{\tilde{\nu}_3'}^2-s_3^2\Delta m_H-s_2^2 c_3^2
 \Delta m_L, \nonumber \\
m_{\tilde{\nu}_2}^2&=&m_{\tilde{\nu}_2'}^2+s_2^2  \Delta m_L, \nonumber \\
B&=&B_{\mu}+s_3^2\Delta B_H-s_2^2 s_3^2 \Delta B_L .
\eea

The additional terms do not significantly effect the vev for the standard
Higgs since the sneutrino vevs and  $\Delta m_{H,L}$, $\Delta B_{H,L}$ 
are suppressed compared to $m_{SUSY}^2$. This remains true even for
$s_3 \sim O(1)$. In addition, we will be considering $s_3 \ll 1 $ to 
account for the atmospheric neutrino scale. Hence, MSSM results are 
hardly altered by additional terms.
The effect of these terms is to induce vevs for the sneutrinos:

\bea \label{omegas}
<\tilde{\nu_2}>&\approx& {c_2s_2s_3\over (m_{\tilde{\nu}_2}^2+D)}
 (v_1\Delta m_{L}-\mu v_2 \Delta B_L)~, \nonumber \\
<\tilde{\nu_3}>&\approx& {c_3s_3\over (m_{\tilde{\nu}_3}^2+D)}
 (v_1(-\Delta m_H+s_2^2\Delta m_L)-\mu v_2 (-\Delta B_H+s_2^2\Delta B_L))~.
\nonumber \\
\eea
We have neglected terms higher order in  $\Delta m_{L,H}, \Delta B_{H,L}$
while writing the above equations. Note that one of the sneutrino field 
($\equiv \tilde{\nu_1}$) does not acquire a vev in this basis. This vev
would arise if Yukawa couplings of the first two generations neglected here
are turned on.

The sneutrino vevs are zero at the boundary scale $\Lambda$ corresponding
to the universal masses. Their weak scale values are determined by solving the
relevant RG equations involving the above differences:
\bea \label{rgq}
{d\;\Delta m_H\over d\;t}&=&3 Y_b(t) ( m_{H_1'}^2+ m_{\tilde{b}}^2
+m_{\tilde{b^c}}^2+ A_b^2)~,   \nonumber\\ 
{d\;\Delta m_L\over d\;t}&=&- Y_{\tau}(t) ( m_{H_1'}^2+ m_{\tilde{\tau}}^2
+m_{\tilde{\tau^c}}^2+ A_{\tau}^2)~, \nonumber\\ 
{d\;\Delta B_H\over d\;t}&=&3 Y_b(t) A_b(t)~,  \nonumber\\
{d\;\Delta B_L\over d\;t}&=&- Y_{\tau}(t)  A_{\tau}~. \eea
Where, $Y_{f}\equiv {\lambda_{f}^2\over (4\pi)^2}$, $m_{\tilde{f}}$
is the mass of the sfermion concerned, $A_f$ are the trilinear soft
SUSY breaking terms and $t=2\;ln(M_{GUT}/Q)$. 

The tree level mass matrix generated due to
these vev can be written in the physical basis $\n_\a$ as:  
\be \label{mtree}
M_0= m_0 O_L \left( 
\ba{ccc}
0&0&0\\
0&s_{\phi}^2& s_{\phi} c_{\phi}\\
0&s_{\phi}c_{\phi}& c_{\phi}^2\\ \ea \right) O_L^T. \ee
Where, 
\be \label{phi}
 \tan \phi= {<\tilde{\nu_2}>\over <\tilde{\nu_3}>} \;\;.\ee
 $O_L$ is defined by eqs.\refs{dia}
and, 
\be \label{m0}
 m_0={\mu (c g^2+g'^2)(<\tilde{\nu}_2>^2+<\tilde{\nu}_3>^2)
 \over 2(-c\m M +2 M_W^2 c_\b  s_\b
(c+ tan^2\theta_W))}. \ee   

\subsection{1-loop mass}

The trilinear interactions in eq.~\refs{w3} lead to  diagrams 
involving squarks and
sleptons in the loop and generate the neutrino masses at the 1-loop
level\cite{loop}. These contributions depend upon the masses as well as
mixing between
the left and the right handed squarks as well as the sleptons. These are
however fixed in terms of the basic parameters of the MSSM. In the present
case, the trilinear couplings are not independent and are controlled
by the fermion masses. As a consequence, the dominant contribution
arises when the b-squark or $\tau$ slepton are exchanged in the loop. 
We shall retain only this contribution.

Let us define:
\bea \label{mix}
 \tilde{b}&=& \tilde{b}_1 \cos \phi_b+\tilde{b}_2 \sin \phi_b~, \nonumber \\
\tilde{b^c}^\dagger&=& \tilde{b}_2 \cos \phi_b-\tilde{b}_1 \sin \phi_b~.
\eea
Where, $\tilde{b}_{1,2}$ are the mass eigenstates with masses
$M_{b_1,b_2}$ respectively. The mixing angles $\phi_{\tau}$ and masses
$M_{\tau_1,\tau_2}$ are defined analogously in case of the tau slepton.

The exchange of b-squark produces the following mass
matrix for the neutrinos:
\be
(M_{1b})_{\a\b}= m_{1b} (O_L)_{\a 3}(O_L)_{\b 3}.
\ee
Due to the antisymmetry of the leptonic couplings in eq.~\refs{w3},
the exchange of the $\t$ slepton leads to the following contribution:
\be \label{mtau}
M_{1\tau}=m_{1\t}\left (\ba{ccc}
O_{L_{13}}^2&O_{L_{13}}O_{L_{23}}&0\\
O_{L_{13}}O_{L_{23}}&O_{L_{23}}^2&0\\
0&0&0\\ \ea \right). \ee
The mixing induced by these contributions is completely fixed
by the matrix $O_L$ while the overall scale of both these contributions 
is set by,
\be \label{mbt}
m_{1b,1\tau}=N_c {m_{b,\t}^3\over 16 \pi^2 v_1^2} \tan\theta_3^2 
\sin\phi_{b,\t}
\cos\phi_{b,\t} \;\;ln \left({M_{b_2,\t_2}^2\over M_{b_1,\t_1}^2}\right).
\ee
Where, $N_c=3,1$ for the $\tilde{b}$ and $\tilde{\tau}$
contribution respectively. The total mass matrix including  the 1-loop
corrections  is given by,
\be \label{total}
{\cal M}_\n=M_0+M_{1b}+M_{1\tau}. \ee

We stress that the above ${\cal M}_\n$ is in the physical basis
with diagonal charged lepton masses. This matrix assumes particularly
simple form when rotated by the matrix $O_L$:
\be \label{masses}
O_L^{T}{\cal M}_\n O_L~ \approx \left ( 
 \ba{ccc}                0&0&0\\
                 0&m_0s_{\phi}^2+m_{1\tau}N_2^{-4}c_2^2s_2^2c_3^2
&m_0s_{\phi}c_{\phi}+m_{1\tau}N_2^{-4}c_2s_2^3c_3^3\\
0&m_0s_{\phi}c_{\phi}+m_{1\tau}N_2^{-4}c_2s_2^3c_3^3&
m_0c_{\phi}^2+m_{1b}+m_{1\tau}N_2^{-4}s_2^4c_3^4\\ \ea
\right) \ee
This explicitly shows that one of the neutrinos is massless in our
approximation of neglecting Yukawa couplings of the first two generations.

The full mixing matrix analogous to the KM matrix is given by,
\be \label {fmix}
 U =O_L O_\n^T.
\ee
Where, $O_\n$ is the matrix diagonalizing the RHS of eq.(\ref{masses}).
As we will show, the mixing angle appearing in $O_\nu$ is small due to
hierarchy in neutrino masses while, the $O_L$ can contain large mixing. 
Hence, the neutrino masses are determined by the matrix (\ref{masses}) and
mixing among neutrinos is essentially fixed by eq.(\ref{ol}). We shall use
these equations in specific case of the gauge mediated models in the next
section.

The above formalism shows that the neutrino masses are greatly
suppressed compared to the typical SUSY breaking scale if
$\Delta m_{H,L},\Delta B_{H,L}$ vanish at some scale $\Lambda$.
The weak scale values of sneutrino vev and hence neutrino masses
follow from evolution of these parameters. It is clear 
from eq.(\ref {rgq}) that the $b$ and $\tau$ Yukawa couplings control the
evolution of sneutrino vev. Similarly, the 1-loop masses following from
eq. (\ref {mbt}) are also controlled by the same couplings. As a result, 
all the effects of lepton number violating parameters $\e_i$ can be rotated away
from
the full Lagrangian when the down quark and the charged lepton Yukawa
couplings vanish. Neutrino masses also vanish in this limit. 
\section{Gauge mediated models and neutrino masses}
The suppression in neutrino masses mentioned above crucially depends upon 
vanishing of  
$\Delta m_{H,L},\Delta B_{H,L}$ at some scale.
This happens in two of the most popular scenarios of supersymmetry
breaking namely, supergravity induced breaking in its minimal form and
the gauge mediated breaking of SUSY \cite{gm1,bkw,gm2}. We now discuss
neutrino masses in
the latter context.

The basic approach adopted in most gauge mediated models of SUSY
breaking is to assume that a singlet sector is responsible for such
breaking. Effect of this is felt by the standard fields through a
messenger sector which is a gauge non-singlet. The minimal version of this 
sector contains a pair of fields $\Phi,\bar{\Phi}$ transforming as
5+$\bar{5}$ representation of the $SU(5)$ group. Their coupling to a 
SUSY breaking field $S$ introduces 
a supersymmetric
mass scale $X\equiv\lambda <S>$ as well as a SUSY breaking (mass)$^2$
differences of order $F_S$. Models with minimal messenger sector are thus
characterized by two parameters $\Lambda\equiv{F_S\over X}$
and $x\equiv{\Lambda \over X}$ with $x\sim O(1)$ on natural grounds.

All the soft parameters related to MSSM fields are fixed at $\Lambda$ in
terms of $\Lambda,x$ and the gauge couplings. The masses of the 
squarks and sleptons and the gauginos are given in this case by:
\bea \label {squ}
m_i^2(X)& =& 2 \Lambda^2 \left\{ C_3 \tilde{\alpha}_3^2(X) + C_2
\tilde{\alpha}^2_2(X) + \frac{3}{5} Y^2 \tilde{\alpha}_1^2(X) 
\right\}f(x), \nonumber \\
M_j(X)&=&\tilde{\alpha}_j(X)~\Lambda~g(x).
\eea
 $m_i^2$ represents the scalar masses with $i$ running over all
 the scalars, whereas, $M_j$ represents the gaugino masses with $j$
 representing the three gauge couplings. The functions $f(x)$ and 
$g(x)$ have been derived in \cite {spnew}. 
Here, 
\be
\tilde{\alpha}_j(X) = {\alpha_j(X) \over (4 \pi)};
\ee
$C_3$ = 4/3,0 for triplets and singlets of $SU(3)_C$, $C_2$ = 3/4,0 for
doublets and singlets of $SU(2)_L$ and Y = Q - $T_3$ is the hypercharge.

In this paper, we will consider two different versions of the model.
The popular \cite{gm1,bkw,r1,r2} minimal messenger model (MMM) which is
further  characterized by the assumption
of the vanishing bilinear ($B$) and trilinear ($A$) soft mass parameters
at $\Lambda$. This is attractive in view of the very restricted structure
it offers. But as we will show it turns out to be too restrictive
if one wants to solve the solar and atmospheric neutrino problems
simultaneously. We shall thus consider an alternative version on
phenomenological grounds in which the  boundary conditions (\ref{squ}) 
are still imposed but the value of $B$ at $\Lambda$ is not taken to be zero. 

In MMM, all the soft parameters in the low energy
theory are essentially determined by one parameter $\Lambda$ since
dependence of the boundary condition eq.( \ref{squ}) on  $x$ is very
mild. In particular, the value of the $B$ parameter at the weak scale
gets fixed through its running. This in turn determines both $\mu$
as well as $\tan\beta$ through the following equations:
\bea \label {mini}
\mu^2&=&{m_{H_1}^2 - m_{H_2}^2 tan^2 \beta \over
 tan ^2 \beta - 1} - {1 \over 2} M_Z^2 ,\nonumber \\
Sin 2 \beta&=&{2 B \mu \over m_{H_2}^2 + m_{H_1}^2 + 2 \mu^2} .
\eea
The presence of the $\e_i$ induces corrections to these equations, but
they are very small as discussed below eq.(\ref{vfull1}).
The eq.(\ref{mini}) therefore holds to a very good approximation.

In spite of the restricted structure, it is possible to self
consistently solve the above equations \cite{bkw,r1,r2} and implement
breaking of the $SU(2)\times U(1)$ symmetry at low energy. Vanishing of
the soft $B$ parameter at $\Lambda$ makes the analysis of this breaking
little more involved than in the case of the supergravity induced breaking.
One needs to include two loop corrections to the evolution of the $B$
parameter and also needs to use fully one loop corrected effective
potential. Details of this analysis are presented in \cite{bkw,r1,r2}. We
follow the treatment given in \cite{r2}. 
The smallness of the $B$ at the weak scale results in this scheme in
relatively large value of $\tan\beta$ and its sign fixes the sign of $\mu$
to be positive. 
The full 1-loop corrected potential was employed in the
analysis of \cite{r2} but it was found that working with RG improved tree
level
potential also gives similar results provided one evolves soft parameters
of the supersymmetric partners up to a scale $Q_0^2\equiv (m_{\tilde{Q}}^2(X) 
m_{\tilde{U}}^2(X))^{1 \over 2}$.
We prefer to follow this approach and use the RG improved tree level
potential of eq.(\ref{vfull}) in order to determine the low energy 
parameters at the minimum.
We have however included  two loop corrections to the RG equations \cite{sp2l}
for $B$ and $\Delta B_{L,H}$ in determining their values at the weak
scale. Use of RG improved tree level potential  allows
us to analytically understand the structure of neutrino masses
and mixing in a transparent way.

The three mass parameters $m_{0,1b,1\tau}$ introduced earlier control the
neutrino masses. $m_0$ is determined by solving RG equations (\ref{rgq})
along with similar ones for parameters occurring in them. We have
numerically solved them imposing eq.(\ref{squ}) as boundary conditions 
at $\Lambda$.
We evolved these equations self consistently  up to the scale
$Q_0$ defined above. The $m_0$ determined in this manner depends upon
$\m$ as well as $\tan\b$ both of which are fixed in terms of $\Lambda$.

The loop contributions are fixed in terms of the squark and slepton masses
and mixings defined in eq. (\ref{mix}). These are determined from the
standard
$2\times 2$ matrices involving left and right squarks and slepton mixing.
The elements in these matrices are also completely fixed in terms of
$\Lambda$. All the three parameters $m_0,m_{1b},m_{1\tau}$ depend upon
an overall scale $s_3$ of the $R$ breaking. For small $s_3^2$ they
roughly scale as $s_3^2$.  The ratios $m_0/s_3^2,~m_{1b}/s_3^2,~
m_{1\tau}/s_3^2$ are thus determined by $\Lambda$ alone \cite{fn2}.
 We have displayed in Fig. 1 variations of
${m_0\over \GeV s_3^2}, {m_{1b}\over m_0}$ and ${m_{1\tau}\over m_0}$ 
with $\Lambda$. One  clearly sees  hierarchy in the loop and sneutrino vev induced
contributions. This hierarchy gets reflected in the neutrino masses and 
one obtains hierarchical neutrino masses independent of the
overall strength of the $R$ violating parameter $s_3$.
The mass ratio and hence the  hierarchy
among neutrino masses are seen to be less sensitive  to $\Lambda$.
The $m_0$ roughly scales linearly with $\Lambda$. But since the over all
scale of $m_0$ is set by $s_3^2$ which is also unknown a change in
$\Lambda$ is equivalent to changing $s_3$. Thus, we may use one specific
value of $\Lambda$ and neutrino mass spectrum is then completely fixed by
three angles $s_{1,2,3}$ or equivalently by the three $R$ violating
parameters $\epsilon_i$.

The $\Delta m_{H,L}, \Delta B_{H,L}$ entering $m_0$ are determined
from the RG equations (\ref{rgq}) and are fixed in terms of $\Lambda$.
For example,
\bea \label{deltas2}
\mu\sim 397.0\GeV~ ,& ~~\tan\b\sim 46.39~ , \nonumber \\
\Delta m_H\sim 192661.23 \GeV^2~, &~\Delta m_L\sim -2392.35 \GeV^2~,\nonumber \\
\Delta B_H\sim -14.07~ \GeV~ ,&~~\Delta B_L\sim 0.12~ \GeV,
\eea
when $\Lambda=100$  \TeV. The suppression in $\Delta m_L,\Delta B_L$ is
due to color factors and larger squark masses compared to the slepton
masses in the model. It follows that the ratio $\tan \phi$ of the
sneutrino vev, eq. (\ref{phi}) gets considerably suppressed even when the
angle $s_2$ is large. We show in Fig 2. the value of $s_{\phi}^2$ as
function of $s_2$ for $\Lambda=100 \;TeV$. Note that this ratio 
is independent of the values of the other $R$ violating parameters
when $s_3$ is small.

The small value of $s_\phi$ leads to very simple expression for neutrino
masses. The neutrino mass matrix in eq.(\ref{masses}) is almost diagonal and one
finds:
\bea \label {teta}
m_{\nu_3}&\sim& m_0+m_{1b}~, \nonumber \\
{m_{\nu_2}\over m_{\nu_3}}&\sim& c_2^2 s_2^2 {m_{1\tau}\over
(m_0+m_{1b})}~\sim 
c_2^2 s_2^2~ (7.1~\times~10^{-3} - 5.6~\times~10^{-3} )~,
\nonumber \\
\theta_{23}^\n&\sim& {m_{1\tau}c_2 s_2^3\over (m_0+m_{1b})}~\sim 
\tan\theta_2 {m_{\nu_2}\over m_{\nu_3}}.
\eea
The masses are fixed in terms of $m_{0,1b,1\tau}$ which are determined 
in terms of $\Lambda$ and $s_3$. The mass ratio is fixed in terms of
$s_2$. The range indicated on the RHS in above equation corresponds to
variation in $\Lambda$ from (51 \TeV - 150 \TeV) and $\theta_{23}^\n$
represents the angle diagonalizing the matrix in eq.(\ref{masses}).

\section{Neutrino masses: Phenomenology}

As discussed in the last section, the model considered here implies
hierarchical masses and large mixing without any fine tuning of the
parameters. We now try to see if the predicted spectrum can be used to
simultaneously reconcile both the solar and the atmospheric neutrino
anomalies. The model is quite constrained. Three neutrino masses and
three mixing angles get completely determined in the model in terms of
four parameters namely, $\Lambda$ and three $R$ violating angles
$s_{1,2,3}$. In particular, the angle $s_1$ characterizing the electron
number violation does not enter the muon and tau neutrino masses, see
eq.(\ref{masses}). The mixing between neutrinos is largely fixed by the matrix
$O_L$ with a small correction coming from the angle $\theta_{23}^\nu$ in
eq.(\ref {teta}). Thus one has approximately,
\bea \label{pattern}
\nu_e&\approx& N_1( c_1\nu_1- s_1 c_2\nu_2 + s_1 s_2c_3 \nu_3) \nonumber\\ 
\nu_{\mu}&\approx& N_1(- s_1 N_2\nu_1+ c_1c_2 N_2^{-1}\nu_2
+c_1 s_2 c_3 N_2^{-1}\nu_3)\nonumber \\
\nu_{\tau}&\approx& N_1^{-1}N_2^{-1} ( -s_2 c_3 \nu_2s_1
 +c_2\nu_3) 
\eea
Note that $s_1$ ($s_2$) determines $\nu_e-\nu_{\mu}\;\; 
(\nu_{\m}-\nu_{\tau})$
mixing. We must thus require $s_2$ to be large in order to account for
the atmospheric muon neutrino deficit. The $s_1$ should be small for the
small angle MSW solution and large for the vacuum oscillation solution to
the solar neutrino problem. As we now demonstrate these constraints are
too tight and one does not obtain parameter space in case of the MMM
allowing simultaneous solution for both these problems.

\subsection{MSW and atmospheric neutrino problem in MMM}

The  angle $s_1$ can be appropriately chosen to fix the required
mixing for the small angle MSW conversion. The angle $s_3$ which
determines the overall scale of neutrino masses is also required to be
small. In such a case,
the survival probability for the atmospheric $\nu_\m$ assumes two
generation form and one can take the restrictions on relevant parameters
from the standard analysis \cite{std}.
We have determined the effective $\nu_{\mu}-\nu_{\tau}$ mixing and 
neutrino masses following from eq.(\ref {fmix}) by the procedure outlined 
in the last section. We show this mixing in Fig.(3a).
 In Fig.(3b), we show the masses 
for two values of $\Lambda = 70 \TeV, 150 \TeV$.
As seen from Fig.(3a), the $s_2 = 0.3 - 0.75 $ leads to the required $\sin^2
2\theta_{\mu\tau} = 0.8 - 1$. Fig.(3b) displays the contours 
corresponding to $\Delta_S\sim( 3. - 12.)~~10^{-6} \eV^2 $ 
and $\Delta_A= (0.3 - 3.)~~10^{-3} \eV^2$
in the $s_2-s_3$ plane. It is seen that  hierarchy among neutrino masses 
obtained in the required region is stronger than needed for a
simultaneous
solution of the solar and atmospheric neutrino problems and there
is no overlapping region in the $s_2-s_3$ plane for a combined solution.
It is of course 
possible to solve each of this problem separately and get the required
amount of mixing as well. 

\subsection{Vacuum oscillations and atmospheric neutrino problem in MMM}

Unlike in the case of the MSW interpretation, the model can nicely account for
the  hierarchies required for solving the solar and atmospheric neutrino
problems through vacuum oscillations. This is displayed in Fig.(4a) where
we show contours corresponding to $\Delta_S = (0.5 - 3)~~ 10^{-10} \eV^2$ and
$\Delta_A = (0.3 - 3)~~ 10^{-3} \eV^2$ in the
$s_2-s_3$ plane. Unlike in case of the MSW conversion, here there is a
large overlap region in $s_2-s_3$ plane which leads to the
 required values for $\Delta_{S,A}$. 
Despite this one unfortunately cannot explain both the problems
simultaneously in a phenomenologically consistent way. This is due to the
very restricted mixing structure displayed in eqs.(\ref{fmix}). The vacuum
oscillation probability in the present case is given by,
\be
P_e=1- 4 U_{e1}^2 U_{e2}^2 \sin^2 \left({\Delta_S t \over 4 E}\right)
-2 U_{e3}^2(1-U_{e3}^2)~, \ee
where the last term comes from the averaged oscillations corresponding to
the atmospheric neutrino scale. Likewise, the muon neutrino survival
probability which determines the atmospheric neutrino flux is given by,
\be
P_\m =1- 4 U_{\m 2}^2 U_{\m 3}^2 \sin^2\left({\Delta_A t \over 4 E}\right) .
\ee
The amplitude of oscillations is controlled by two effective
angles:
\be \sin^2 2\theta_S=4 U_{e1}^2 U_{e2}^2~,\;\;\;\;
     \sin^2 2\theta_A=4 U_{\m2}^2 U_{\m 3}^2 .\ee
The matrix U appearing above is given by eq.(\ref{fmix}).
Restrictions on these angles  required for a combined solution of the
solar and atmospheric anomaly are worked out in \cite{osland} for
different values of $U_{e3}$. Independent of the 
chosen values for $U_{e 3}$ one requires,
\be \label {rest}
\sin^2 2\theta_S = 0.5 - 1~,\;\;\;\; \sin^2 2\theta_A = 0.8 - 1~.\ee
It is possible to choose these angles independently and satisfy above
equations in a generic three generation case. In our case, the 
mixings are also determined in terms of $s_{1,2}$ through eq.(\ref{ol}). We have
plotted the contours corresponding to restrictions in eq.(\ref{rest}) 
in Fig. 4b.
It is seen that there is no region in $s_1-s_2$ plane for which the solar 
and vacuum mixing angles can be simultaneously large ruling out the
  possibility of reconciling atmospheric anomaly with vacuum 
solution in the case of the MMM.

\section{Non-minimal model and neutrino anomalies}

We had restricted our analysis so far to the MMM which is characterized
by eq.(\ref{squ}) and the vanishing of the $B$ and $A$ parameters at $\Lambda$.
Apart from  predictivity, there are no strong theoretical
arguments in favour of this minimal choice. 
One could consider variations of the MMM which in general result in 
introduction of additional low energy parameters. A class of non-minimal
models could contain more complicated messenger sector which would influence
boundary conditions in eq.~(\ref{squ}). Alternatively, one may keep the 
same messenger sector but introduce some direct coupling between messenger
and matter fields. This could result in non-zero B values at $\Lambda$. 
In fact, B
gets generated \cite{dine,pomerol} in models which  try to understand origin
of $\mu$ term in gauge mediated scenario \cite{murayama}. B may be 
generated in the absence of messenger-matter coupling if MSSM itself is
extended.

We shall not consider any specific model here, but would adopt a
purely phenomenological attitude to point out possible ways which can
allow simultaneous understanding of the solar and atmospheric neutrino
anomalies. It turns out that prediction of $m_0$ is quite sensitive
to the sign of $\mu$ term which is fixed to be positive in MMM. This 
follows from eq.(\ref{omegas}) which shows that two contributions
 to $\left< \tilde{\nu} \right>$ add or cancel depending on the sign of
$\mu$. We may thus consider a slightly less 
restrictive form of the MMM in which we regard the value and sign of $B_\m$  
as independent parameters to be determined phenomenologically. 
We still assume that mechanism
responsible for generation of $B$ parameters does not distinguish between 
leptonic and the Higgs doublet $H_1$ and hence 
$B_\m$ and $B_i$ coincide at the scale $\Lambda$. Due to this, sneutrino
vevs are still characterized by the differences in eq.(\ref{deltas1}) 
and hence are suppressed. The boundary conditions on soft
masses are still assumed to be given by eq.(\ref{squ}).
This particular scenario is now
characterized by  parameters $\Lambda$ and $B_\m$. As follows from the
minimum equation, (\ref{mini}), one may regard the value of $\tan\beta$
and sign of $\mu$ as independent parameters instead of $B_\m$.
The magnitude of $\m$ is determined in terms of these parameters 
by eq.(\ref{mini}).  It is now possible to simultaneously account for the
 atmospheric neutrino deficit and have the MSW conversion for the
 solar neutrinos.  This is depicted in Fig.~(5) in which
we show contours (in $s_2-s_3$ plane) corresponding to
$\Delta_S= 3 - 12 ~~ 10^{-6} \eV^2,\Delta_A= 0.3 - 3 ~~10^{-3} \eV^2$ for
negative $\mu$ and two representative
values of $\tan\beta=40,50$. The magnitude of $\mu$ gets fixed 
by eq.~(\ref{mini}) to 379.28 ~\GeV~and 382.03 ~\GeV~in the respective cases.
 It is seen that
now there is a considerable overlap where two mass scales arise
simultaneously. As mentioned before, these masses are independent of
the value of $s_1$ which can be chosen in the required range namely,
$$s_1 = 0.0225 - 0.071 $$ to allow MSW conversion.
 The angle $\theta_A$
relevant for the atmospheric anomaly coincides roughly with $s_2$ and
as follows from Fig.(5), one can simultaneously account for mixing as well
as masses needed to solve the atmospheric and the solar neutrino
problems. 

\section{Discussion}

The supersymmetric standard model  contains natural source of lepton
number violation and hence of neutrino masses. The resulting neutrino
  mass pattern is quite constrained if source of lepton number
 violation is provided by soft bilinear operators and 
if the SUSY breaking is introduced through gauge mediated interactions. 
This scenario has the virtue that
one can obtain hierarchical masses and large mixing in the neutrino sector.
The hierarchy in masses results from hierarchy in the two different sources
of neutrino masses while large mixing can be linked to ratio of
$R$ violating parameters $\epsilon_i$. 
Overall scale of neutrino mass is  set by $s_3$ and by
the Yukawa couplings of the $b$ and $\tau$ . Neutrino masses
are thus naturally suppressed and hierarchical. One however needs to
assume relatively suppressed $R$ violation, i.e. $s_3\sim 10^{-3}$
in order to obtain the mass scale relevant for the atmospheric neutrino
anomaly. This requires that $\e_2\sim \e_3\sim 1\GeV$ when $\m\sim 1
\TeV$.

In case of the minimal gauge mediated model, the three neutrino
masses and three mixing angles are 
controlled by four  parameters $\Lambda$ and $s_{1,2,3}$. This proves to
be quite constraining and does not allow one to obtain simultaneous
solution of the solar and atmospheric neutrino anomalies. However, 
a non-minimal version which allows negative $\mu$ parameter is capable of
accommodating the MSW effect and atmospheric neutrino anomaly. The number
of parameters needed are still less than in the models based 
on the minimal supergravity scenario. 

Although we concentrated on the bilinear terms a similar situation is
obtained if neutrino masses are generated from the trilinear lepton number
violating terms also. Here also, neutrinos obtain their masses from
the loop as well as sneutrino vev induced contribution \cite{sr}. Unlike
the bilinear terms, the number of possible
trilinear terms is quite large and one needs to fine tune some of 
these or impose additional discrete symmetries in order to get a 
 consistent picture \cite{manuel}.

We have mainly concentrated here on generating the atmospheric 
and the solar scales. The LSND result or the presence of hot neutrino
dark matter needs an additional scale. It should be possible to introduce
this through a sterile neutrino. Supersymmetry with gauge mediated
interaction may be ideally suited to do so as it may explain both the
lightness as well as the very existence of a sterile state \cite{sterile}.

\noindent
{\bf Acknowledgments:} We have greatly benefited from numerous discussions
we had with K. S. Babu, G. Rajasekaran, Probir Roy and J. W. F. Valle.

\newpage
\begin{figure}[h]
\epsfxsize 15 cm
\epsfysize 20 cm
\epsfbox[25 151 585 704]{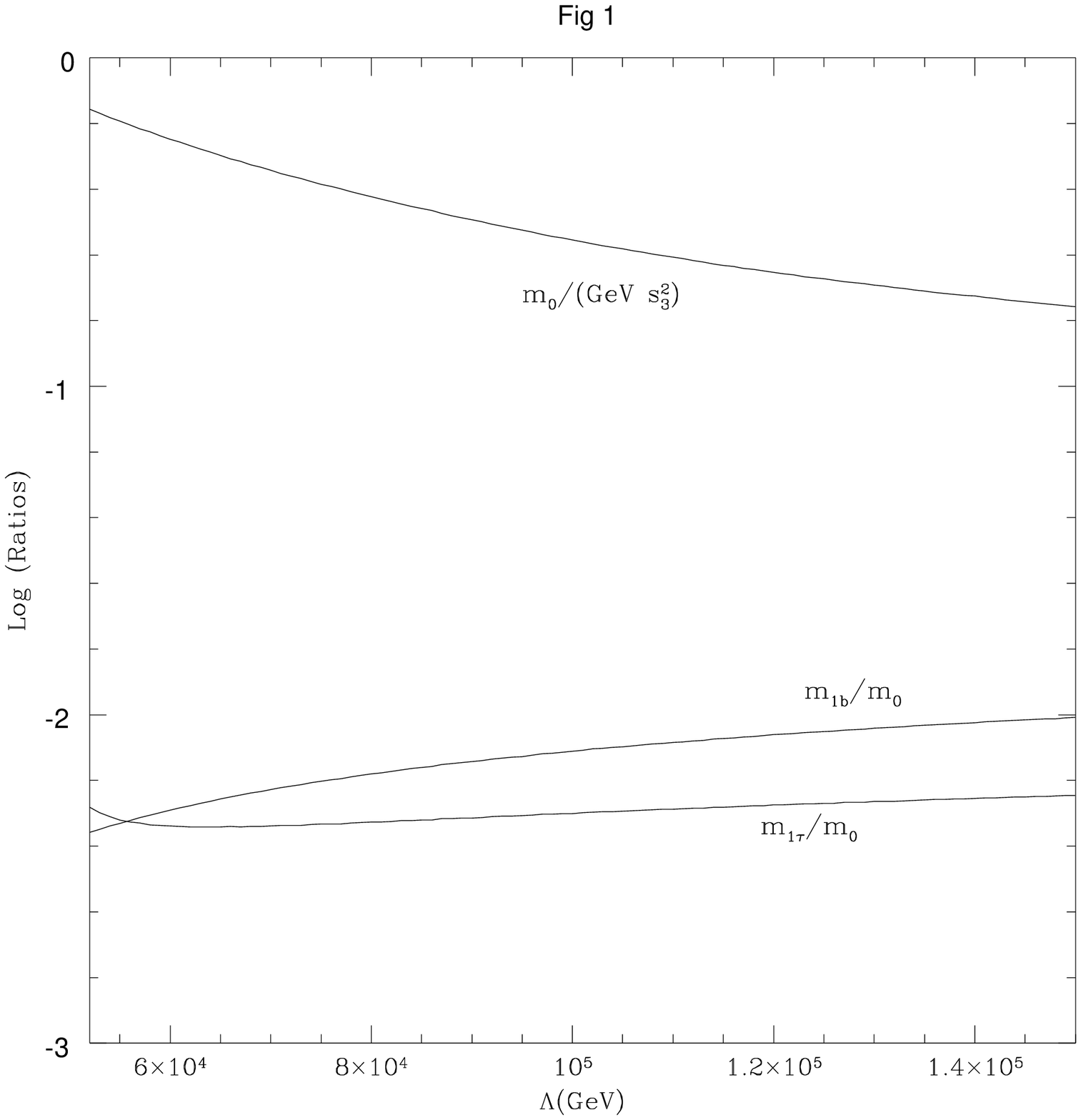}
\end{figure}
\noindent
{\bf Fig. 1}. ~\sl{ The variations of ${m_0 \over (\GeV s_3^2)}$,
${m_{1b} \over m_0}$,${m_{1 \tau} \over m_0}$ are shown here with
respect to $\Lambda$. $m_0$ mildly depends upon $s_2$ and the displayed
curve is for $s_2 = 0.8$.}
\newpage
\begin{figure}[h]
\epsfxsize 15 cm
\epsfysize 20 cm
\epsfbox[25 151 585 704]{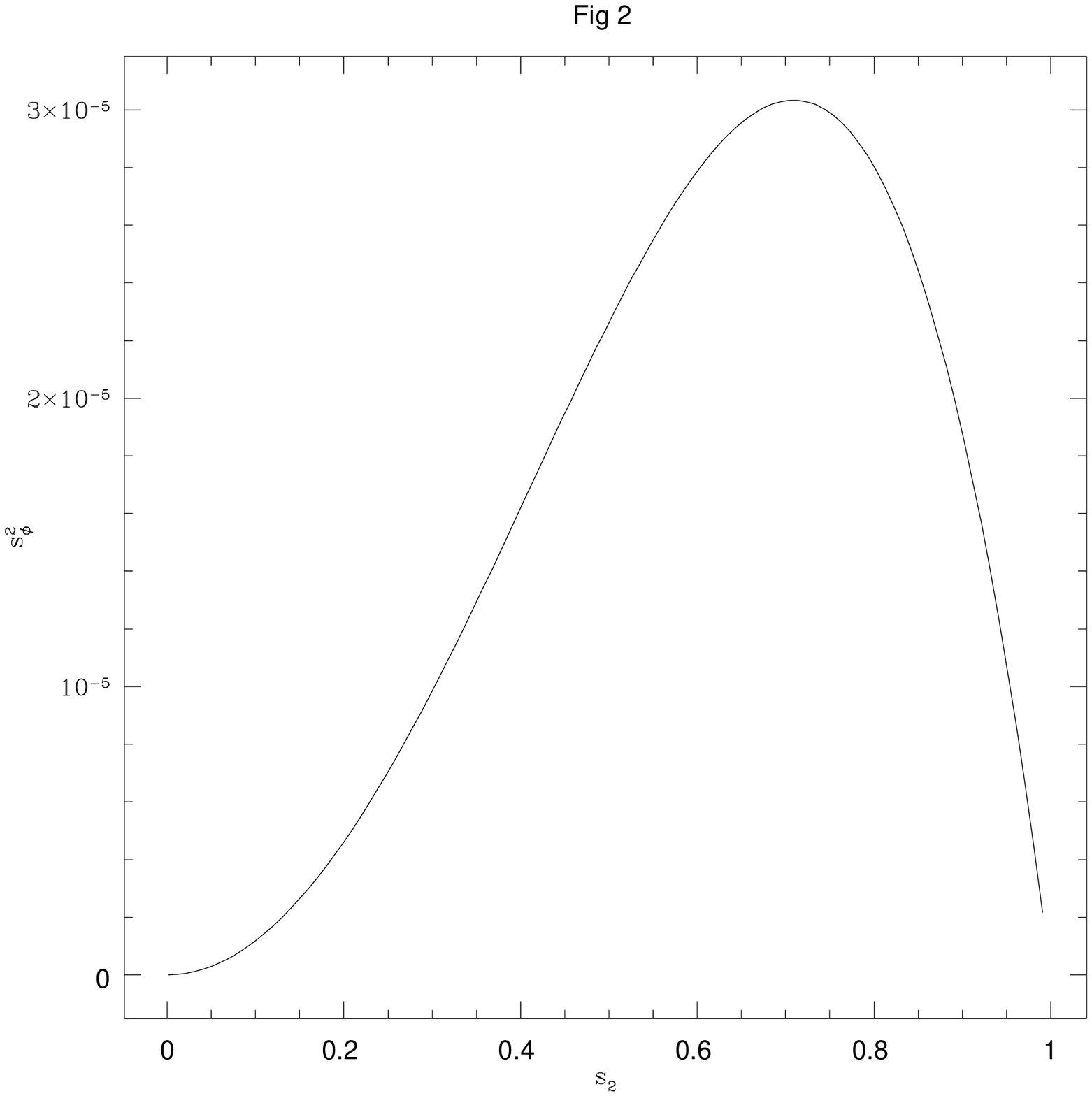}
\end{figure}
\noindent
{\bf Fig. 2}. ~\sl{ The function $s_{\phi}^2$ is plotted here with respect
to $s_2$.}

\newpage
\begin{figure}[h]
\epsfxsize 15 cm
\epsfysize 20 cm
\epsfbox[25 151 585 704]{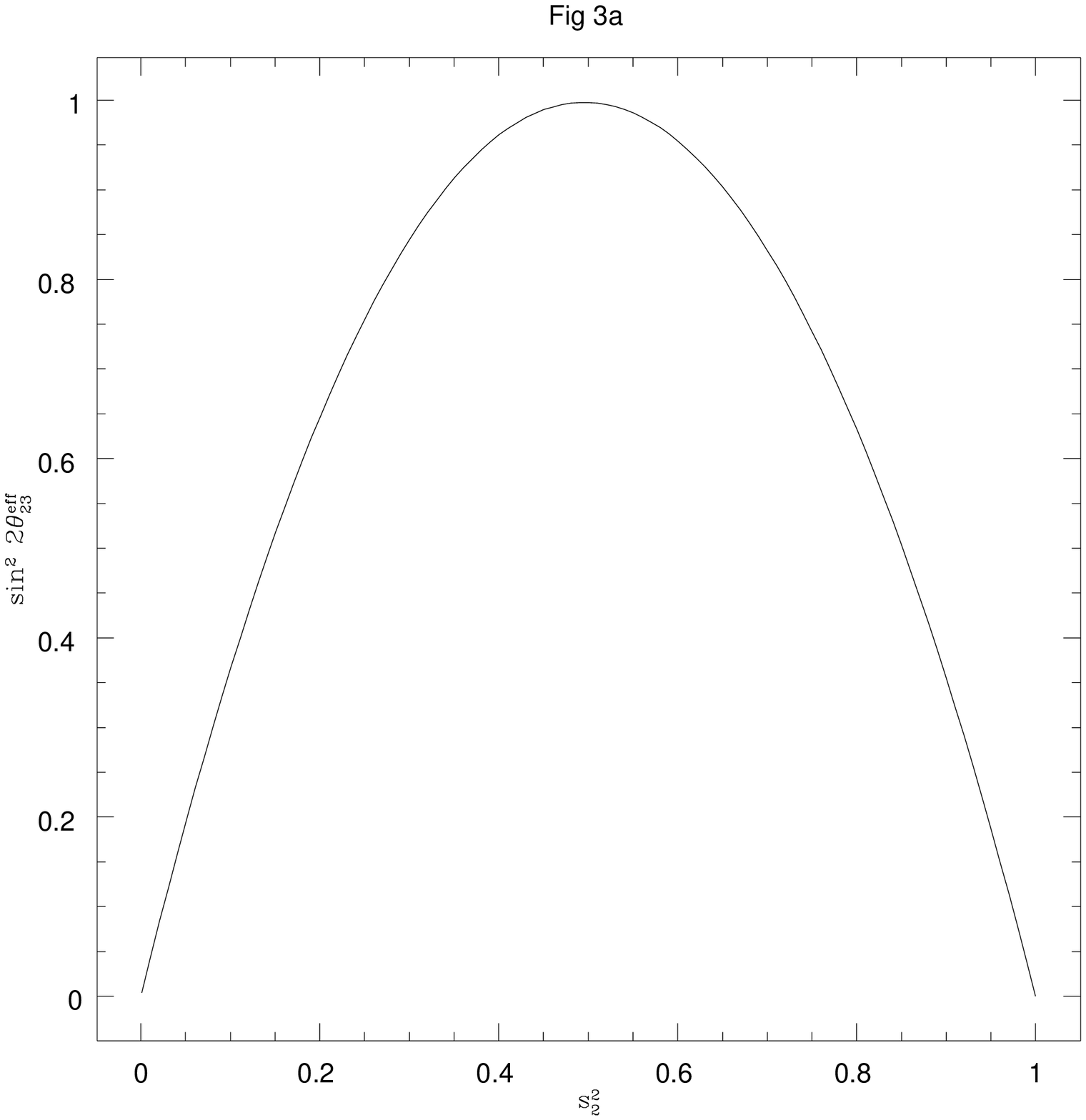}
\end{figure}
\noindent
{\bf Fig. 3a}. ~\sl{ The effective $\nu_\mu $-$\nu_\tau$  mixing angle
is plotted here with respect to $s_2^2$ for $\Lambda = 100 \TeV$
in the case of minimal messenger model.}
\newpage
\begin{figure}[h]
\epsfxsize 15 cm
\epsfysize 20 cm
\epsfbox[25 151 585 704]{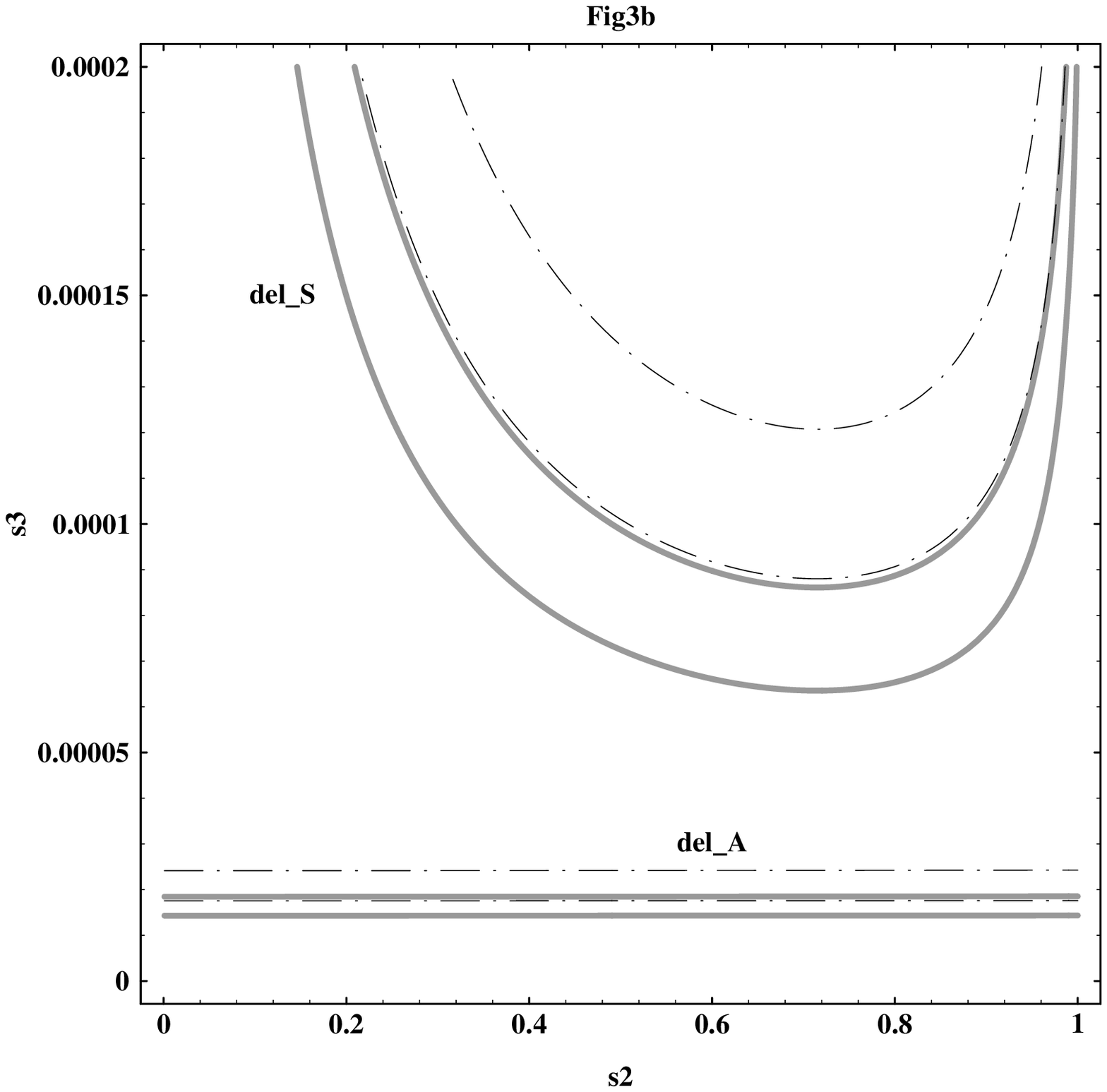}
\end{figure}
\noindent
{\bf Fig. 3b}. ~\sl{ Contours of $\Delta m^2$ are plotted in {\bf MMM} case, 
for $\Lambda$ = 70 \TeV~(continuous lines) and $\Lambda$ = 150 \TeV ~(dash-dot).
 For $\Delta_{A}$, the upper (lower) lines correspond to $3~\times~10^{-3}
 \eV^2$ ($ 0.3~\times~10^{-3} \eV^2 $). For $\Delta_{S}$, the upper(lower) lines
 correspond to $12~\times~10^{-6} \eV^2$ ($3~\times~10^{-6} \eV^2$)}.

\newpage
\begin{figure}[h]
\epsfxsize 15 cm
\epsfysize 20 cm
\epsfbox[25 151 585 704]{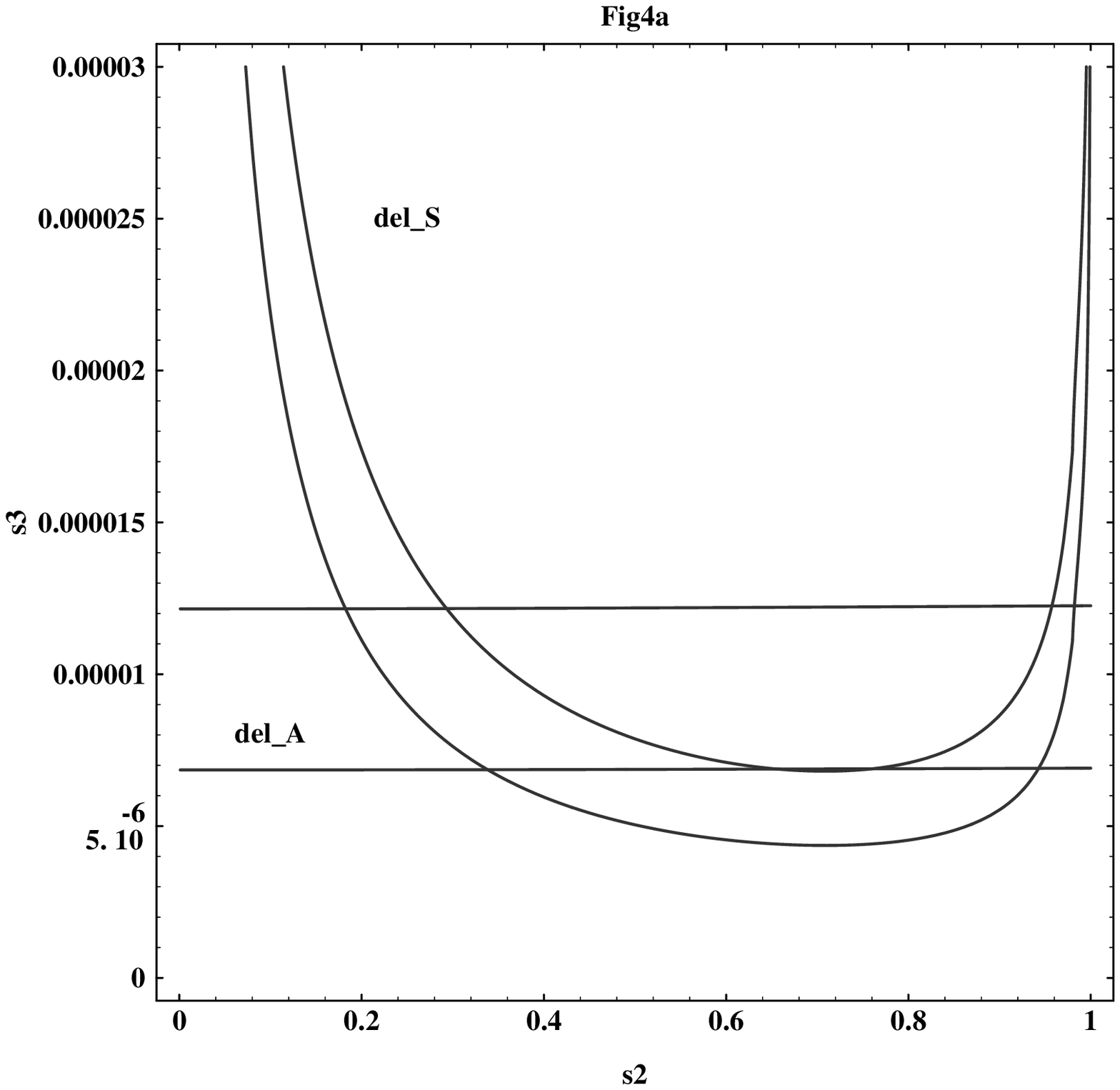}
\end{figure}
\noindent
{\bf Fig. 4a}. ~\sl{ Contours of $\Delta m^2$ in MMM are plotted 
 for $\Lambda$ = 100 \TeV. For $\Delta_{A}$, the upper (lower) line corresponds
 to $3~\times~10^{-3} \eV^2$ ($ 0.3~\times~10^{-3} \eV^2 $). For $\Delta_{S}$, 
the upper (lower) line corresponds to $3~\times~10^{-10} \eV^2$
 ($0.5~\times~10^{-10} \eV^2$).}

\newpage
\begin{figure}[h]
\epsfxsize 15 cm
\epsfysize 20 cm
\epsfbox[25 151 585 704]{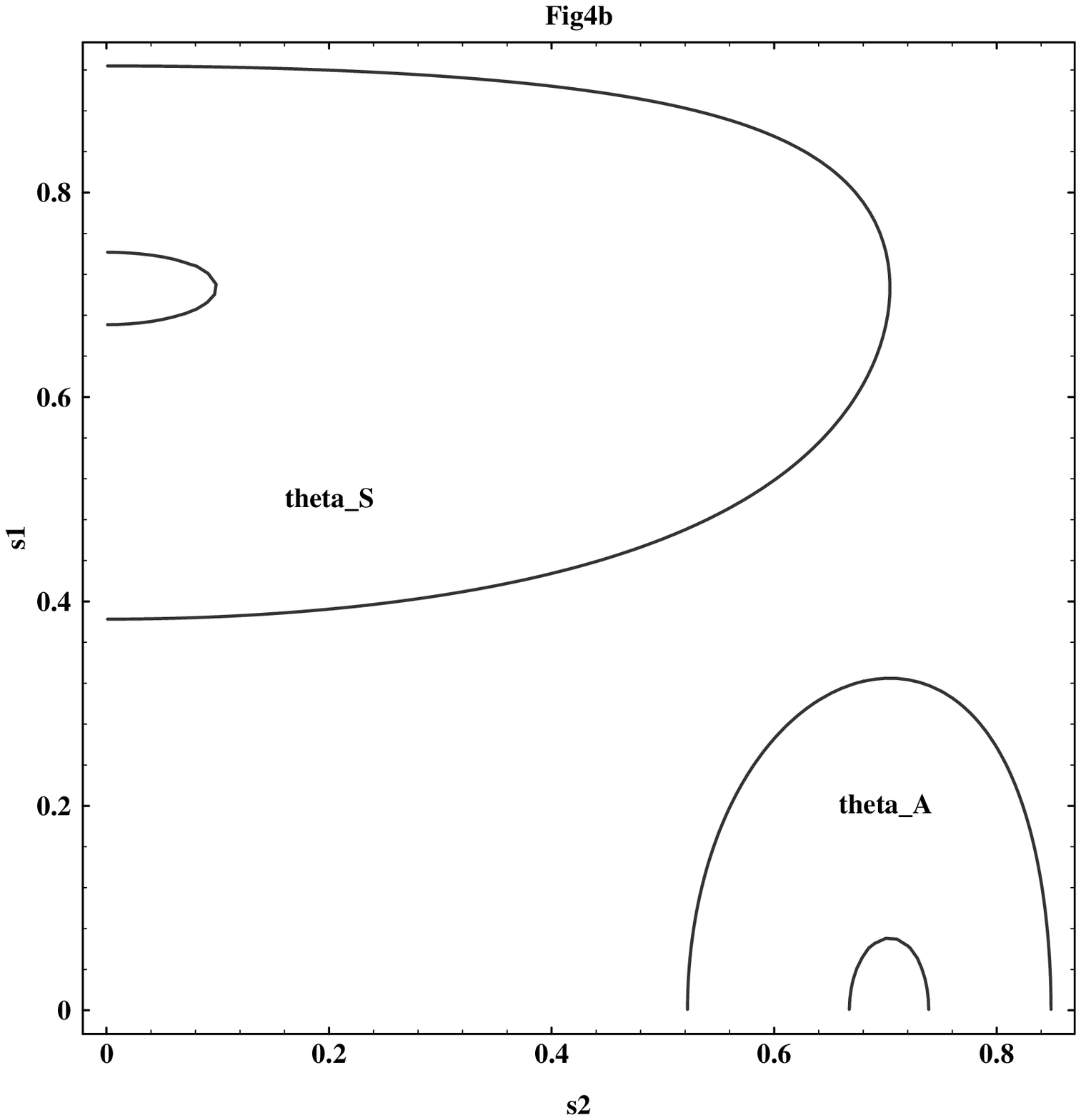}
\end{figure}
\noindent
{\bf Fig. 4b}. ~\sl{The effective  $Sin ^2  2 \theta_{S}$ and 
$Sin ^2 2 \theta_{A}$ are plotted in the case of the minimal messenger model.
 The inner lines represent contours for 0.9 in both the cases whereas, 
the outer lines correspond to contours for 0.5 (0.8) for $Sin^2 2 \theta_{S}$ 
($Sin^2 2 \theta_{A}$).}

\newpage
\begin{figure}[h]
\epsfxsize 15 cm
\epsfysize 20 cm
\epsfbox[25 151 585 704]{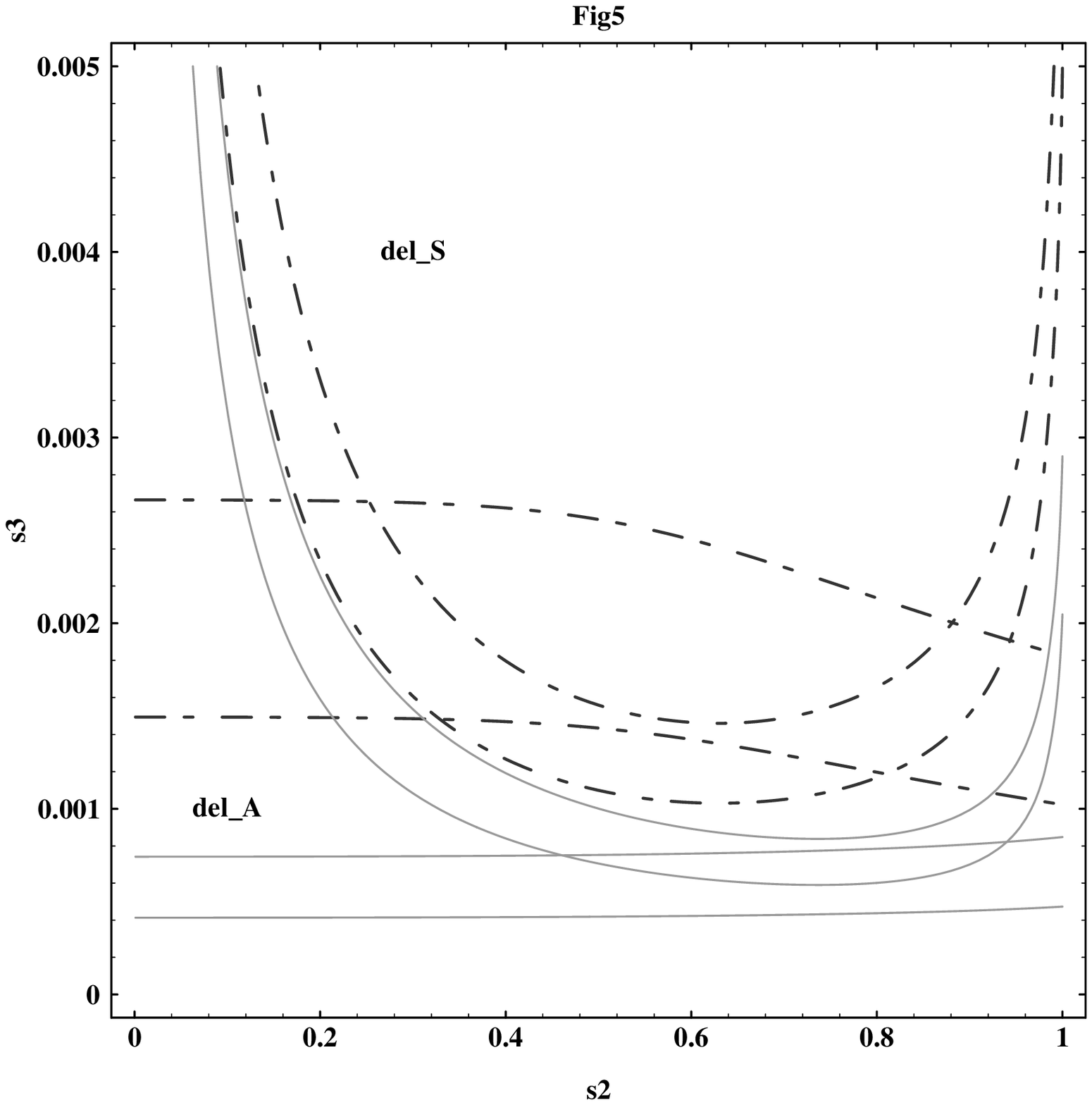}
\end{figure}
\noindent
{\bf Fig. 5}. ~\sl{ Contours of $\Delta m^2$ are plotted in {\bf Non- MMM} case
with $\mu < 0$, for tan $\beta$ = 50 ~(continuous lines) and
 tan $\beta$ = 40 ~ (dash-dot) with $\Lambda = 100 \TeV$. 
 For $\Delta_{A}$, the upper (lower) lines correspond to $3~\times~10^{-3}
 \eV^2$ ($ 0.3~\times~10^{-3} \eV^2 $). For $\Delta_{S}$, the upper 
(lower) lines correspond to $12~\times~10^{-6} \eV^2$ ($3~\times~10^{-6}
 \eV^2$)} .


\begin{thebibliography}{99}
\bibitem{fn1}We use the notation $H'_1 H_2 \equiv \epsilon_{ab} 
H_1^{'a} H_2^{b}$ with $\epsilon_{12} = -\epsilon_{21} = 1$.
\bibitem{f2} Note that we have neglected here a sub-dominant contribution
to ${\cal M}_l$ arising due to sneutrino vevs.
\bibitem{fn2}$m_0$ also depends on $s_2$ very mildly through the
vev $\left< \tilde{\nu}_2 \right>$ which is greatly suppressed compared
to $\left< \tilde{\nu}_3 \right>$.
\end{thebibliography}
\end{document}